\documentclass[aps,prd,twocolumn,showpacs,amsmath]{revtex4-1}
\usepackage{graphicx}
\usepackage{amssymb}

\newcommand{\bra}[1]{\langle #1 | \,}
\newcommand{\ket}[1]{\, | #1 \rangle}

\newcommand{\be}{\begin{equation}}
\newcommand{\ee}{\end{equation}}
\newcommand{\bea}{\begin{eqnarray}}
\newcommand{\eea}{\end{eqnarray}}
\newcommand{\besa}{\begin{subeqnarray}}
\newcommand{\eesa}{\end{subeqnarray}}
\newcommand{\bean}{\begin{eqnarray*}}
\newcommand{\eean}{\end{eqnarray*}}

\newcommand{\mbr}{\mathbf{r}}
\newcommand{\mbk}{\mathbf{k}}

\begin{document}

\title{Universal properties of Fermi gases in arbitrary dimensions}

\author{Manuel Valiente}
\affiliation{Lundbeck Foundation Theoretical Center for Quantum System Research, Department of Physics and Astronomy, Aarhus University, DK-8000 Aarhus C, Denmark}
\author{Nikolaj T. Zinner}
\affiliation{Lundbeck Foundation Theoretical Center for Quantum System Research, Department of Physics and Astronomy, Aarhus University, DK-8000 Aarhus C, Denmark}
\author{Klaus M{\o}lmer}
\affiliation{Lundbeck Foundation Theoretical Center for Quantum System Research, Department of Physics and Astronomy, Aarhus University, DK-8000 Aarhus C, Denmark}

\date{\today}

\begin{abstract}
We consider spin-$1/2$ Fermi gases in arbitrary, integer or non-integer spatial dimensions, interacting via a Dirac delta potential. We first generalize the method of Tan's distributions and implement short-range boundary conditions to arbitrary dimension and we obtain a set of universal relations for the Fermi gas. Three-dimensional scattering under very general conditions of transversal confinement is described by an effectively reduced-dimensional scattering length, which we show depends on the three-dimensional scattering length in a universal way. Our formula for non-integer dimensions interpolates between the known results in integer dimensions 1, 2 and 3. Without any need to solve the associated multichannel scattering problem, we find that confinement-induced resonances occur in all dimensions different from $D=2$, while reduced-dimensional contacts, related to the tails of the momentum distributions, are connected to the three-dimensional contact by a correction factor of purely geometric origin.

\end{abstract}

\pacs{03.75.Ss, 
      03.65.Nk,  
      67.85.Lm, 
}

\maketitle

\section{Introduction}
The dependence of physical properties on dimensionality is an important
issue with great theoretical and practical ramifications. The most well-known 
examples are arguably theories containing extra dimensions, such 
as the original attempt of Nordstr{\"o}m, Kaluza, and Klein to unify gravity 
and electrodynamics by adding an extra spatial dimension \cite{kaluza}. 
These ideas are currently a backbone in modern thereotical high-energy frameworks 
\cite{schwarz,add,rs,dgp}.
However, the quest to go in the other direction, i.e. to systems with spatial
dimension less than three, is equally important since
low-dimensional structures such as wires and surfaces dominate our daily
surroundings. In fact, quantum dynamics in one- and two-dimensional systems can be 
very different from the three-dimensional case \cite{giamarchi}.

An obvious mathematical question related to variation in dimensionality 
concerns interpolation between integer values, and its relevance
in physics. This notion, however, turns out to have great value in such areas as 
the characterization of chaotic phenomena \cite{strogatz}, in 
surface growth \cite{surface}, on the study of phase transitions \cite{zinn}, 
for domain formation in spin models \cite{spinmodels}, and also for 
quantum computation \cite{markham} on fractal lattices \cite{gefen}.
From a more formal theoretical point of view, variation of dimensionality
beyond integer has had tremendous impact on particle physics, statistical 
physics, and condensed-matter through the famous $\epsilon$-expansion by
Wilson \cite{wilson} and dimensional regularization by t'Hooft \cite{thooft}.


The advantage of studying cold atomic gases is the clean conditions that 
allow an isolation of specific effects of interest \cite{BlochReview}. The 
experimental control over the interactions means that one has direct
access to the strongly-coupled regime where the 
scattering amplitude in a two-component Fermi gas saturates the 
unitarity bound of quantum mechanics (or equivalently when the $s$-wave
scattering length diverges). An experimental surge has managed to map out
important features of the universal thermodynamics of this system \cite{ufermi}.
However, as shown by Tan \cite{Tanenergetics} (see also \cite{OPE}), the 
unitary regime has a direct connection between microscopic quantities such as
the momentum distribution and the macroscopic thermodynamic observables.
These relations have subsequently been demonstrated in experiments on 
three-dimensional gases \cite{Stewart,kuhnle}. In lower dimensions, 
similar microscopic relations have been 
derived \cite{werner2010,VZM,barth2011,hofmann2012,braaten2012,combescot2009}, 
and experiments testing these relations in two dimensions are 
on-going \cite{vogt2012}.

In the current paper we provide an interpolation/extrapolation 
formula for Tan's contact relations as a function of the 
spatial dimension, $D$, which is considered a continuous 
variable in the general setup that we discuss. The objective is
to provide interpolation formulas for the contact identities 
that can be used for instance in interpolating from known 
limits into unknown territory, and testing the results of different
analytical and numerical approaches to strongly-coupled dynamics. 
This is similar in spirit to the $\epsilon$-expansion based on the 
properties of the Fermi gas
in $D=2$ and $D=4$ interpolated to $D=3$ and compared to 
Monte Carlo calculations by Nishida and Son \cite{nish2006,nish2007}.
These relations should be useful in general strong-coupling 
investigations, for instance when applying the renormalization 
group using the $\epsilon$-expansion. More generally, they may
provide hints of whether certain confining geometries 
allow access to systems where the dynamics indicate an 
effective non-integer dimensionality. An example could be a 
holographic model where the boundary turns out to have fractal 
dimension, as for instance seen in the spin models mentioned above. 

Using these universal relations we find that, under very general conditions, effective interactions in quasi-$D$-dimensional Fermi gases, which appear as emergent theories for transversely confined three-dimensional systems, are related to the original three-dimensional interaction through a universal functional form. This new type of emergent universality has a strong predictive power: the so-called confinement-induced-resonances \cite{Olshanii} are shown to occur in all dimensions except for the two-dimensional case, which agrees with the known results in integer dimensions \cite{Olshanii,PetrovShlyapnikov}. 

The paper is structured in the following way. In Sec. \ref{sectiontwo}, we introduce Tan's $\eta$-distribution in terms of integrals over $D$-dimensional momentum space. In Sec. \ref{sectionthree}, we present the corresponding set of Tan relations, providing connections between a number of microscopic and macroscopic properties of the Fermi gas in $D$ dimensions. In Sec. \ref{sectionfour}, we recall Olshanii's and Petrov-Shlyapnikov's results for scattering confined to one and two dimensions, respectively, and we show that they are special results of $D$-dimensional expressions, which link the scattering lengths and contacts to their three-dimensional equivalents. Sec. \ref{sectionfive} concludes the manuscript.


\section{Formalism}\label{sectiontwo}
We consider a spin-$1/2$ Fermi gas with contact $s$-wave two-body interactions in arbitrary spatial dimension $D$. The scattering of a pair of fermions is then completely characterized by a single quantity, namely the scattering length $a_D$.
Contact interactions have to be regularized in $D\ge 2$ and renormalized (in any dimension) to fit the corresponding $D$-dimensional scattering amplitude for two-particles or equivalently, when a bound state exists, to the two-body binding energy in vacuum. This procedure can be done in a number of ways, all equivalent to each other: non-perturbative Gell-Mann--Low renormalization \cite{GellmannLow,Adhikari}, dimensional regularization \cite{tHooft,KaplanReno}, $D$-dimensional Bethe-Peierls boundary conditions \cite{BethePeierls}, Fermi-Huang pseudopotential \cite{Huang} or its $\Lambda$-family \cite{OlshaniiPricoupenko} and the more recent method of Tan's distributions \cite{Tanenergetics,Tanarxiv,ValienteTan,ValienteHard,VZM}. In this work, we will make use of the latter method.

In order to treat non-integer-dimensional spaces and, in particular, integrals of functions of $D$-dimensional variables, we first need to define Euclidean space in arbitrary dimension. There are several ways of doing this, e.g. Stillinger \cite{Stillinger} and Wilson spaces \cite{WilsonSpaces}. The spaces differ in the fact that Stillinger spaces do not have vectorial character, while Wilson spaces do. Integration in these two spaces can be successfully defined, although the appearance of negative integration weights seems unavoidable. For all purposes of this paper, both Stillinger and Wilson spaces are equivalent, and we will only need simple integrals where no conceptual problems arise. In particular, for (hyper-)spherically symmetric integrands we can safely use the replacement:
\be
\int d^{D}x \to \lambda(D) \int_{0}^{\infty} dx x^{D-1},
\ee
where $\lambda(D)=2\pi^{D/2}/\Gamma(D/2)$ is the $D$-dimensional solid angle.

We now derive Tan's $\eta$ distribution \cite{Tanenergetics} in arbitrary non-even dimension $D$. Note that the case $D=4$ is not relevant --- there is no scattering by a Dirac delta even after renormalization --- while for $D=2$ we derived the corresponding expression in \cite{VZM}. The $\eta$ distribution is defined via $\eta(k)=1$ ($k<\infty$) and
\be
\int d^Dk \eta(k)\frac{1}{\hbar^2k^2/m+E_B}=0,\label{etacondition}
\ee
where $E_B$ is given by \cite{Jensen}
\be
E_B=\frac{4\hbar^2}{m a_D^2}\left[\frac{\Gamma^2(D/2)\sin(\pi(D-2)/2)}{\pi(D-2)/2}\right]^{2/(D-2)},
\ee
and coincides with the two-body binding energy if the scattering length $a_D$ is positive. The $\eta$-distribution, as defined in Eq. (\ref{etacondition}), provides a simple way of implementing short-range boundary conditions for a $D$-dimensional delta interaction and its use is completely equivalent to cut-off and dimensional regularization-renormalization procedures \cite{ValienteTan}.

From previous analyses for dimensions $D=1,2,3$ \cite{ValienteHard,VZM,ValienteTan}, we see that
\be
\eta(k)=1-\mathcal{G}(k)\delta(1/k),\label{Eqone}
\ee
where $\mathcal{G}(k)$ is determined from Eq. (\ref{etacondition}) and is therefore given by
\be
\mathcal{G}(k)=\frac{\hbar^2}{mk^{D-1}} \int_{0}^k dq \frac{q^{D-1}}{\hbar^2q^2/m+E_B}.\label{Eqtwo}
\ee
The series representation of the above integral for non-even dimensions is obtained from
\begin{align}
&\int_{0}^{k} dq\frac{q^{D-1}}{\frac{\hbar^2 q^2}{m}+E_B} = \frac{1}{E_B}\left(\frac{mE_B}{\hbar^2}\right)^{D/2}\times \nonumber \\
&\left[\frac{\pi}{2}\csc(\pi D/2) + \frac{k^D}{D} \left(\frac{\hbar^2}{mE_B}\right)^{D/2}\frac{2\Gamma(1+D/2)}{\Gamma(D/2)}\mathcal{S}(D;k)\right],\label{Eqthree}
\end{align} 
where
\be
S(D;k)=\sum_{n=1}^{\infty} \frac{(-1)^{n+1}}{(D-2n)\left(\frac{\hbar^2}{mE_B}\right)^{n}k^{2n}}.\label{sumS}
\ee
We note that terms higher than $n=\lfloor D/2 \rfloor$ in the series (\ref{sumS}) do not contribute to the integral in Eq. (\ref{etacondition}).

\section{Tan relations in $D$ dimensions}\label{sectionthree}
We show here the set of universal relations for Fermi gases in arbitrary dimensions. We single out the two-dimensional case, for obvious reasons, which has been derived previously using Tan's distributions \cite{VZM}. Proceeding in a way which parallels refs. \cite{VZM,Tanenergetics}, we obtain the complete set of Tan relations.

\paragraph{Energy theorem}
The energy $E_D$ satisfies the following energy theorem
\be
E_D=\sum_{\mbk,\sigma} \eta(k) \frac{\hbar^2k^2}{2m} n_{\mbk,\sigma}+\langle W \rangle,
\ee
where $W$ is a single particle trapping potential acting on each fermion in the gas and $n_{\mbk,\sigma}$ is the occupation number
of the momentum state with momentum $\mbk$ and spin $\sigma$. Using Eqs. (\ref{Eqone},\ref{Eqtwo},\ref{Eqthree},\ref{sumS}), we can expand the above relation to obtain
\begin{align}
E_D&=\frac{\lambda(D)V_D}{(2\pi)^D}\frac{\hbar^2}{2m}\sum_{\sigma}\lim_{\Lambda\to \infty}\left[\int_{k<\Lambda}d^Dk k^{2}\frac{n_{\mbk,\sigma}}{\lambda(D)}\right.  \nonumber\\
&\left.-\left(\frac{mE_B}{\hbar^2}\right)^{(D-2)/2}\frac{\pi}{2}\csc(\pi D/2) C_D \right.\nonumber \\
&\left.-\frac{\Lambda^{D+4}}{D}\left(\frac{\hbar^2}{mE_B}\right)^{D/2}\frac{2\Gamma(1+D/2)}{\gamma(D/2)}\mathcal{S}(D,\Lambda)n_{\Lambda,\sigma}\right]\nonumber \\
&+\langle W \rangle,
\end{align}
where $V_D$ is the $D$-dimensional hypervolume and $C_D=\lim_{\mbk\to \infty}\mbk^4 n_{\mbk,\sigma}$ is called the $D$-dimensional contact.

\paragraph{Adiabatic relation}
We obtain
\be
\frac{dE_D}{da_D}=\frac{\alpha_D C_D}{a_D^{D-1}},\label{adiabatic}
\ee
where
\begin{align}
\alpha_D&=(D-2)\left(\frac{\hbar^2}{m}\right)^{2-D/2}\frac{V_D\lambda(D)\pi}{2(2\pi)^D}\times \nonumber \\
&\csc(\pi D/2)(a_{D}^{2}E_B)^{D/2-1}.\label{alpha}
\end{align}

\paragraph{Generalized virial theorem}
Given a trapping potential of the form $W\propto r_D^{\beta}$, we obtain
\be
E_D=\frac{\beta+2}{2}\langle W \rangle-\frac{\alpha_D C_D}{a_D^{D-2}}.
\ee

\paragraph{Pressure relation}
If the system is homogeneous, that is, $W\equiv 0$, the pressure relation reads 
\be
P_DV_D = \frac{1}{D}\left[2E_D + \frac{\alpha_D C_D}{a_D^{D-2}}\right].
\ee

\paragraph{Two-particle loss rate}
Adding a small imaginary part to the scattering length, that is, replacing $a_D$ with $a_D+ia_I$, we obtain the last of the relations for the two-body loss rate $\Gamma_D$
\be
\Gamma_D = -2a_I \frac{\alpha_D C_D}{a_D^{D-1}} + O(a_I^2).
\ee

\section{Effective $D$-dimensional interactions}\label{sectionfour}
The effective reduction of a three-dimensional zero-range interacting two-body system placed under the influence of a $(3-D)$-dimensional trap to an integer lower dimension ($D=1,2$) is achieved using a method due to Olshanii \cite{Olshanii} or variations thereof \cite{PetrovShlyapnikov,Drummond1,ValienteMolmer}. For the effective one-dimensional case under two-dimensional isotropic harmonic confinement the reduced one-dimensional scattering length is given by \cite{Olshanii}
\be
a_1 = -\frac{\ell_0^2}{a_3}+\frac{\chi}{\sqrt{2}}\ell_0,\label{D1}
\ee
where $\ell_0=\sqrt{\hbar/m\omega}$ is the oscillator length, $\chi=\lim_{s\to \infty} (\int_{0}^s dtt^{-1/2}-\sum_{s'=1}^{s}s'^{-1/2})\approx 1.46$. In two dimensions, under a one-dimensional transversal harmonic trap, the effective scattering length reads \cite{PetrovShlyapnikov}
\be
a_2=2e^{-\gamma_E}\sqrt{\frac{\pi}{B}}\ell_0 \exp(-\sqrt{\pi/2}\ell_0/a_3),\label{D2}
\ee
where $\gamma_E$ is Euler's number and $B\approx 1$.  

Actually, dimensional reduction can be done from any $D'\ne 3$ to a lower dimension $D$, but we will assume that the reference space has dimension three. The general idea also applies to non-integer $D$ and involves the calculation of zero-energy (up to a constant offset) wave functions $\Psi$ that behave as
\be
\Psi(\mbr)\to \psi(\mbr_D)\phi_0(\mbr_{3-D}),
\ee
as ${r_D\to \infty}$, where $\mbr_D$ represents the $D$-dimensional relative position, $\psi(\mbr_D)$ is the $D$-dimensional zero-energy wave function corresponding to an effective scattering length $a_D$ (to be calculated), and $\phi_0$ represents the ground-state mode of the $(3-D)$-dimensional external trap. 

This kind of effective dimensional reduction is a multichannel scattering problem with an infinite number of coupled channels and, obviously, requires regularization and renormalization due to the three-dimensional zero-range interaction. The multichannel problem needs to be solved, in principle, in order to obtain the functional form of the $D$-dimensional scattering length $a_D$ in terms of the original three-dimensional one $a_{3}$ in vacuum, which can be regarded as the renormalization condition.     

We here derive the general functional form for $a_D\equiv F(a_3,D)$ using only dimensional analysis and the adiabatic theorem, Eq. (\ref{adiabatic}), as follows. We first need a length scale, which we denote by $\bar{x}_D$ which, in the case of an isotropic harmonic confinement must be proportional to the oscillator length. The $D$-dimensional contact and energies are denoted by $C_D$ and $E_D$, respectively. The number of identical reduced-dimensional subsystems embeded in the three-dimensional space is denoted by $\mathcal{N}_D$ (typically this is realized through confinement in optical lattice potentials \cite{BlochReview} such that in $D=1$ each of these systems is a tube while in $D=2$ they are pancakes). Assuming that each of these subsystems is well separated from the rest and has energy $E_D$, the total energy of the system $E_3$ will be given by $E_3=\mathcal{N}_D E_D$, up to an $a_3$-independent energy offset. The three-dimensional volume of each subsystem is given by its hypervolume $V_D$ times its transversal extent $\bar{x}_D^{3-D}$, so the total volume of the system is given by $V_3 = \mathcal{N}_D\bar{x}_D^{3-D}V_D$. Using these facts in the three- and $D$-dimensional adiabatic relations, Eq. (\ref{adiabatic}), we obtain
\be
\mathcal{N}_D\frac{\alpha_DC_D}{[F(a_3,D)]^{D-1}}\frac{\partial F}{\partial a_3}(a_3,D) = \frac{\hbar^2 \bar{x}_D^{3-D}\mathcal{N}_D V_D}{4\pi m a_3^2}C_3.\label{equti}
\ee
From Eqs. (\ref{D1}) and (\ref{D2}), we see that $F^{1-D}\partial F/\partial a_3\propto a_3^{-2}$ in one and two dimensions. Assuming that this is also the case in arbitrary dimension $D$, we arrive at the condition
\be
\frac{1}{[F(a_3,D)]^{D-1}}\frac{\partial F}{\partial a_3}(a_3,D) = \frac{\kappa_D}{a_{3}^2},\label{difeq}
\ee
where $\kappa_D$ is independent of $a_3$. From Eq. (\ref{difeq}) and for $D\ne 2$ we obtain
\be
a_D=F[a_3,D]=\left[\frac{1}{(D-2)(\gamma_D+\kappa_D/a_3)}\right]^{1/(D-2)},\label{aDeff}
\ee
where $\gamma_D$ is a constant. In two dimensions we obtain
\be
a_2=F[a_3,2]=e^{\gamma_2}e^{-\kappa_2/a_3}.\label{a2eff}
\ee 
The above expressions (\ref{aDeff}) and (\ref{a2eff}) are the desired relations between effective $D$-dimensional scattering lengths and the three-dimensional scattering length in vacuum. For all dimensions $D$ there are two {\it a priori} unknown constants $\gamma_D$ and $\kappa_D$. To obtain the constant $\gamma_D$ we need to solve the full multichannel scattering problem as discussed above, but $\kappa_D$ can be exactly calculated from a mean-field shift in the coupling constant, as shown in the Appendix. Note that, as must obviously be, Eq. (\ref{aDeff}) for $D=1$ and Eq. (\ref{a2eff}) agree with the known results (\ref{D1}) and (\ref{D2}), respectively. The case $D=3$ is trivially satisfied, with $\kappa_3=1$ and $\gamma_3=0$.

The above analysis has some striking, non-trivial consequences. First, one obtains a confinement-induced resonance (CIR) corresponding to $a_D=0$ for all $D<2$ whenever $a_3=-\kappa_D/\gamma_D$. For all dimensions $2<D<3$ this matching condition results in $a_D\to \infty$, and it is only the two-dimensional case where no resonant behavior may occur. Second, while the values of $\kappa_D$ and $\gamma_D$ are model-dependent, relations (\ref{aDeff}) and (\ref{a2eff}) are universal. These are valid for any isotropic transversal $(3-D)$-dimensional confinement potential provided that either (i) $(3-D)$-dimensional center-of-mass and relative coordinates can be separated \cite{Saenz} (as is the case for harmonic confinement) or (ii) the corresponding CIR is so broad as compared to further resonances that a so-called single pole approximation \cite{ValienteMolmer} for the effective scattering length is valid, which is typically the case \cite{Saenz,ValienteMolmer}. 

If we use now Eqs. (\ref{D1}) and (\ref{D2}) in Eq. (\ref{equti}), in the case of isotropic harmonic confinement we find the proportionality constants that relate the effective contacts in $D=1,2$, to the three-dimensional one:
\begin{align}
C_2&=\frac{1}{\sqrt{2\pi}}\frac{\bar{x}_2}{\ell_0}C_3 \\
C_1&=\frac{1}{2\pi}\left(\frac{\bar{x}_1}{\ell_0}\right)^2C_3.
\end{align}
For general dimension, we can obtain the above relation up to a multiplicative constant and it reads
\be
C_D = \frac{\hbar^2 \bar{x}_D^{3-D}}{4\pi m \xi_D\kappa_D} C_3\equiv \Delta_D C_3,\label{generaleq}
\ee
where $\xi_D=\alpha_D/V_D$. Since we have $\xi_D\propto \hbar^2/m$, we have from dimensional grounds that $\kappa_D \propto \ell_0^{3-D}$ and $\Delta_D$ is dimensionless and $a_3$-independent. Therefore, $\Delta_D$ is of purely geometric origin and we call it the $D$-dimensional geometric correction factor (GCF).    

Some important remarks about the GCF are in order. We have assumed that the system is tightly confined by a $(3-D)$ dimensional trap, which implies that the effective $D$-dimensional Fermi energy is much smaller than the first excited energy ($\propto \hbar \omega$) of the transversal trap, $E_F^{(D)}\ll \hbar \omega$. The GCF is in this case a constant, but the $D$-dimensional and three-dimensional contacts are not necessarily the same. To see this, we can estimate the two- and one-dimensional transversal extents $\bar{x}_{2}$ and $\bar{x}_1^2$, respectively, for the case of harmonic confinement. We define $\langle r_D^2\rangle=\bra{\psi_0}r_D^2\ket{\psi_0}$, where $\ket{\psi_0}$ represents the single-particle ground-state in the transversal trap. Then, the quasi-two-dimensional system will have a extent given by $\bar{x}_{2}=2\sqrt{\langle r_1^2\rangle}=\sqrt{2}\ell_0$, which gives $\Delta_2=1/\sqrt{\pi}$. In the one-dimensional case, assuming a cylindrical shape for the quasi-one-dimensional tubes, we have $\bar{x}_1^2=\pi \langle r_2^2 \rangle = \pi \ell_0^2$, which gives $\Delta_1=1/2$. These correction factors must be taken into account when relating $C_3$ to the $D$-dimensional contact $C_D$, in order to verify universality in effectively reduced-dimensional Fermi gases.

In a recent experiment concerning CIRs \cite{Innsbruck}, anharmonicities in the trapping potential are present, which are not described by our previous analysis. This implies that close to anharmonic CIRs \cite{Saenz,Drummond}, the quasi-$D$-dimensional contact $C_D$ is not proportional to the three-dimensional contact $C_3$. In the quasi-1D isotropic case we may, on the other hand, use a two-pole approach -- see ref. \cite{ValienteMolmer} for the single-pole approximation -- to describe the presence of a second resonance, which can be justified in the spirit of Pad{\'e} approximants. We assume that the quasi-1D interaction strength $g_1=-2\hbar^2/ma_1$ has the form
\be
-g_1\approx \frac{1}{\Gamma_1+\Omega_1a_3}+\frac{1}{\Gamma_2+\Omega_2a_3},
\ee
where $\Gamma_i$ and $\Omega_i$ ($i=1,2$) mark the positions $a_3^{(i)}=-\Gamma_i/\Omega_i$ of the two CIRs, while $\Omega_1/\Omega_2$ defines the relative width of the resonances when they are well separated ($|a_3^{(1)}-a_3^{(2)}|\gg |\Omega_i^{-1}|$ for $i=1,2$). We then have, close to the broad CIR (e.g. $a_3\approx a_3^{(1)}$), from Eq. (\ref{difeq}),
\be
\frac{C_3}{C_1}\propto \frac{\Gamma_1^2}{\Omega_1}+2\Gamma_1\left(a_3+\frac{\Gamma_1}{\Omega_1}\right)\left(\frac{\Gamma_1\Omega_1}{\Gamma_1\Omega_2-\Gamma_2\Omega_1}-1\right)\label{anharmonic}
\ee
The above expression gives the linear correction to the relation between 3D and quasi-1D contacts in the isotropic, anharmonic case. The first term on the r.h.s. of Eq. (\ref{anharmonic}) corresponds to the proportionality constant in the harmonic case, while the linear term reflects slight deviations due to anharmonicity in the vicinity of the broad CIR. 

\section{Conclusions}\label{sectionfive}
We have derived universal Tan relations for spin-$1/2$ Fermi gases 
interacting via a Dirac delta potential in arbitrary integer and 
non-integer spatial dimensions. The formalism introduced was examplified by considering extrapolation
of known results for the three-dimensional case where the  
contact of the unitary Fermi gas has been calculated for the ground-state
energy using the $\epsilon$-expansions \cite{nish2006,nish2007}. 
Starting from this three-dimensional initial condition, we studied the properties of 
Fermi gases 
subject to transversal $(3-D)$-dimensional confinement  which 
results in an effective realization of $D$-dimensional Fermi gases. Using 
the universal relations, we found that, in the limit of tight transversal 
confinement, the effective $D$-dimensional two-body interactions exhibit a 
universal functional dependence on the three-dimensional scattering length 
in vacuum. This universal function implies the appearance of confinement-induced 
resonances \cite{Olshanii,PetrovShlyapnikov} in any integer or non-integer dimension, 
with the only exception being 
the well-known two-dimensional case. Furthermore, the quasi-$D$-dimensional 
contact is related to the three-dimensional one, by a 
correction factor whose value is dictated by the geometry of the transversal trap.
The example can be generalized to the case of a
$D_0$-dimensional reference space and $(D_0-D)$-dimensional confinement, such as is often 
seen in higher-dimensional theories ($D_0>3$) where our everyday observable Universe 
is embedded as a submanifold. The prescription provided here shows how to 
extrapolate higher-dimensional universal quantities down to the relevant submanifold.

Tan relations are of relevance 
for the $\epsilon$-expansion \cite{wilson} around two and four dimensions, and 
can be of use in other setups where dimensional interpolation or extrapolation
of physical quantities is necessary. A case of interest could be 
extradimensional or holographic theories where higher-dimensional observables
must be projected onto lower dimensional physical realization of the 
predicted dynamics. These theories are used in addressing 
strongly-coupled dynamics, which is a limit in which Tan relations 
have proven particularly powerful in recent cold atomic gas experiments. 
We hope that the current work can provide a foundation for exploring the 
implications of Tan relations within other fields as well.

\acknowledgments
M.V. acknowledges support from a Villum Kann Rassmussen Block Scholarship. N.T.Z. is supported by the Danish Council for Independent Research | Natural Sciences.

\appendix

\section{Calculation of $\kappa_D$ through the mean-field shift in the coupling constant}
In general, the effective $D$-dimensional coupling constant $g_D$ is related to the scattering length via
\be
g_D=(D-2)\lambda(D)(\hbar^2/m)a_D^{D-2}.\label{coupling}
\ee
When the system is transversely confined by a $(3-D)$-dimensional trap, it is given by \cite{Olshanii}
\be
g_D=g_3|\phi_0(\mathbf{r}_{3-D}=0)|^2+\mathcal{O}(g_3^2),\label{gdg3}
\ee
to lowest order in $g_3$, where $\phi_0$ is the normalized ground-state wave function of the $(3-D)$-dimensional trapped system in the relative coordinate. 

For the case of a harmonic trap, we obtain \cite{Lohe} 
\be
|\phi_0(\mathbf{r}_{3-D}=0)|^2 = \frac{1}{(2\pi)^{(3-D)/2}\ell_0^{3-D}}.
\ee
Equation (\ref{gdg3}) becomes in this case 
\be
g_D=\frac{g_3}{(2\pi\ell_0^2)^{(3-D)/2}}+\mathcal{O}(g_3^2).
\ee
Using the corresponding expression for $g_D$ in terms of $a_D$, we find
\be
a_D=\left[\frac{4\pi}{(D-2)\lambda(D)(2\pi\ell_0^2)^{(3-D)/2}}a_3\right]^{1/(D-2)},
\ee
to lowest order. From Eq. (\ref{aDeff}) to lowest order in $a_3$ we then obtain
\be
\kappa_D=\frac{\sqrt{\pi}}{2^{(D-1)/2}\Gamma(D/2)}\ell_0^{3-D}.
\ee

\end{document}